\documentclass[english,aps,prl,superscriptaddress,twocolumn]{revtex4}
\usepackage[latin1]{inputenc}
\usepackage{graphicx}
\usepackage{amssymb}

\usepackage{bm}
\usepackage{amsfonts}

\makeatother

\usepackage{babel}
\makeatother

\begin{document}

\title{Adiabatic dynamics in open quantum critical many-body systems}

\author{Dario Patan\`e}
\affiliation{Departemento de Física de Materiales, Universitad Complutense, $28040$
  Madrid, Spain}
\affiliation{MATIS-INFM $\&$ Dipartimento di Metodologie Fisiche e Chimiche (DMFCI),
  Universit\`a di Catania, viale A. Doria 6, 95125 Catania, Italy}
\author{Alessandro Silva}
\affiliation{The Abdus Salam International Centre for Theoretical Physics, Strada
  Costiera $11$, $34100$ Trieste, Italy }
\author{Luigi Amico}
\affiliation{Departemento de Física de Materiales, Universitad Complutense, $28040$
  Madrid, Spain}
\affiliation{MATIS-INFM $\&$ Dipartimento di Metodologie Fisiche e Chimiche (DMFCI),
  Universit\`a di Catania, viale A. Doria 6, 95125 Catania, Italy}
\author{Rosario Fazio}
\affiliation{International School for Advanced Studies (SISSA), Via Beirut $2-4$,
  $34014$ Trieste, Italy }
\affiliation{NEST-CNR-INFM $\&$ Scuola Normale Superiore Piazza dei Cavalieri
  7, I-56126 Pisa, Italy }
\author{Giuseppe E. Santoro}
\affiliation{International School for Advanced Studies (SISSA), Via Beirut $2-4$,
  $34014$ Trieste, Italy }
\affiliation{CNR-INFM Democritos National Simulation Center, Via Beirut $2-4$,
  $34014$ Trieste, Italy }
\affiliation{The Abdus Salam International Centre for Theoretical Physics, Strada
  Costiera $11$, $34100$ Trieste, Italy }

\begin{abstract}
The purpose of this work is to understand the effect of an external
environment on the adiabatic dynamics of a quantum critical system.
By means of scaling arguments we derive a general expression for
the density of excitations produced in the quench as a function of its
velocity and of the temperature of the bath. We corroborate
the scaling analysis by  explicitly solving the case of a one-dimensional quantum Ising model coupled to an Ohmic bath.
\end{abstract}

\maketitle

The understanding of the non-equilibrium dynamics of strongly
correlated quantum systems is one of the most challenging problems
of modern condensed matter physics. Interest on this subject has
been revived recently by unprecedented experimental breakthroughs in
the context of cold atomic gases (see e.g~\cite{exps}).
Non-equilibrium conditions in cold gases can be realized
controllably in various ways, e.g. by a proper choice of the initial
state, or by changing the Hamiltonian in time. In view of the variety
of situations that can be studied in this context, it is of
paramount importance to find paradigmatic situations that allow the
 general features of the non-equilibrium dynamics of
many-body systems to be understood. One such paradigm is obtained
when the parameters of a quantum system close to a quantum phase
transition are varied in time in such a way as to traverse the
quantum critical point. Because of the vanishing of the gap $\Delta$
at criticality, a finite density of defects is generated, no matter
how slow is the quench, as first shown in
Ref.~\cite{zurek05,polkovnikov05}. The density of defects is a {\em
universal} scaling function of the quench velocity $v$, as in the
Kibble-Zurek (KZ) mechanism~\cite{KZ} originally derived for
classical continuous phase transition. In addition to its intrinsic
interest, this problem is relevant to adiabatic quantum
computation~\cite{farhi01} and quantum annealing~\cite{santoro}.

The intense theoretical activity following
Refs.~\cite{zurek05,polkovnikov05} has clarified several important
issues (see~\cite{kzquantum} and references therein) on the
adiabatic, phase coherent dynamics of closed many-body critical
systems. Closed systems, however, are only
idealizations: any quantum system is weakly coupled to an environment
inducing relaxation and dephasing. This observation motivated a
series of recent studies, in particular  on  the effect of
classical~\cite{Fubini07} and quantum~\cite{Mostame07} noise acting
uniformly on a quantum Ising chain, and on the effect of local noise
on a disordered Ising chain with up to 20 spins~\cite{Amin08}.
However, the most natural and important question remained
answered: {\it to what extent is it possible to describe universally the production of
defects in an adiabatic quench in the presence of
dissipation and dephasing} ? In this Letter, we answer this
question by showing with a general scaling analysis that {\em even in
the presence of an external environment} the adiabatic dynamics of
open critical systems is governed by {\em universal scaling laws}
(with modified exponents). To support this statement, we solve by
means of quantum kinetic equations the adiabatic dynamics of a
quantum Ising chain coupled to a local external environment, leading
to the relaxation of all quasiparticle modes.

%
\begin{figure}
\includegraphics[scale=0.35]{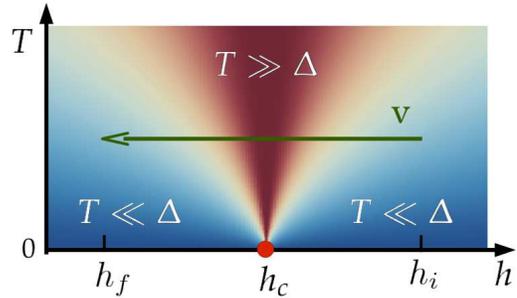}
\caption{\label{fig:fig1} A sketch of the crossover diagram at
finite temperatures due to the presence of the quantum critical point, where the gap $\Delta\to 0$.
The system traverses the quantum critical region in a time $t_{QC}$.}
\label{fig1}
\end{figure}

\paragraph*{Scaling analysis of defect formation.}
%
At equilibrium, all the characteristic features of a quantum phase
transition leave an imprint at low temperatures, leading, close to
the quantum critical point, to a crossover at temperatures
$T\sim\Delta$. For $T\ll\Delta$, the low frequency dynamics can be
described semiclassically in terms of quasi-particle excitations,
while in the \emph{quantum critical region} (red region in
Fig.\ref{fig1}) $T\gg\Delta$ and quasi-particle excitations no
longer exist~\cite{SachdevBOOK}.  The purpose of the analysis below
is to discuss the universal fingerprints left on the nonequilibrium
dynamics by the quantum phase transition in the presence of a bath.
For this sake, we now start by presenting a scaling analysis
describing the influence of a bath at temperature $T$ on the
production of defects.

We start by considering a linear quench of the control
parameter $h$ from an initial value $h_{i}$ to a final one $h_{f}$
across the critical point $h_{c}$ at a speed $v$. The system is
initially in equilibrium with the bath, whose temperature $T$ is
kept constant during the quench. This is sketched in Fig.\ref{fig1}.
The passage through the quantum critical region ($T\gg\Delta$) leads
to substantial heating effects because of the relatively high bath
temperature with respect to the characteristic energy scale $\Delta$
for the system excitations. We shall see that, as long as this is
the most important time interval for the bath-systems interaction,
only the low energy details of the system spectrum matter, and
universality holds.

The universality of the defect production in the presence of a bath
follows from two conditions, discussed below. First of all, the
density of excitations $\mathcal{E}$ can be written as
\begin{equation} \label{eq:ebath+eclean}
\mathcal{E} \simeq \mathcal{E}_{KZ}+\mathcal{E}_{inc}
\end{equation}
where $\mathcal{E}_{KZ}$ is the coherent contribution, present also
in a closed system, and $\mathcal{E}_{inc}$ is the incoherent one
due to the presence of the bath. For a closed system, the density of
excitations was found to scale as $\mathcal{E}_{KZ}\propto
v^{d\nu/(z\nu+1)}$~\cite{zurek05,polkovnikov05}, where $d$ is the
spatial dimension of the system and $\nu,\ z$ are the correlation
length and the dynamical critical exponents respectively.
%
%
Moreover we  assume that the bath does not influence
the system in the semiclassical regions ($T\ll\Delta$).
Hence, we consider thermal excitations  predominantly created inside
the quantum critical region, where $T\gg \Delta$, at a rate
$\tau^{-1}\propto\alpha T^{\theta}$, $\alpha$ being the system-bath
coupling constant. The dynamics for the population of the (excited)
mode $k$ is thus governed by the rate equation
$\frac{d}{dt}P_k=-\tau^{-1}(P_k-P_k^{th}\left(h_{c}\right))$,
where $P_{k}^{th}(h_c)$ is the critical thermal equilibrium distribution
 inside the quantum critical region.
Integrating this rate equation we obtain $P_{k}(h_{f})\sim(1-e^{-\tau^{-1}t_{QC}})P_{k}^{th}\left(h_{c}\right)$,
where $t_{QC}=2T^{1/\nu z}v^{-1}$ is the time spent going through the quantum critical region
(see Fig.\ref{fig1}), whose boundaries are given by $T\sim\Delta\sim|h-h_{c}|^{\nu z}$.
Integrating over all $k$-modes we obtain:
\begin{equation} \label{eq:ebath_approx_general}
\mathcal{E}_{inc} \propto
\left(1-e^{-\frac{t_{QC}}{\tau}}\right)\int dE\ E^{\frac{d}{z}-1}P_{k}^{th}\left(h_{c}\right) \;,
\end{equation}
where we used the scaling of the critical energy $E\propto k^{z}$.
Finally, since the thermal distribution $P_{k}^{th}$ is a function of $E/T$,
changing variable to $E/T$ leads to
\begin{equation} \label{eq:Ebath_generic_QPT}
\mathcal{E}_{inc}\propto\alpha v^{-1}T^{\theta+\frac{d\nu+1}{\nu z}} \;,
\end{equation}
valid in the limit $T^{1/\nu z}\ll v\tau$.
The  $v^{-1}$ scaling of $\mathcal{E}_{inc}$ is directly related to
the time spent inside the quantum critical region.
 The crossover from the coherent to the incoherent defect production
is reached when $\mathcal{E}_{inc}\simeq\mathcal{E}_{KZ}$, giving
\begin{equation} \label{eq:vbath_generic_QPT}
v_{cross}\propto\alpha^{\frac{\nu z+1}{\nu(z+d)+1}}
T^{\left(1+\frac{(\theta-1)\nu z}{\nu(z+d)+1}\right)\left(1+\frac{1}{\nu z}\right)}\;.
\end{equation}
The different scaling with $v$ of $\mathcal{E}_{KZ}$ and
$\mathcal{E}_{inc}$ implies that for fast quenches, $v>v_{cross}$,
the KZ contribution dominates, while for slower quenches,
$v<v_{cross}$, the incoherent contribution due to the thermalization
induced by the bath is the most important.

\begin{figure}
\includegraphics[scale=0.32]{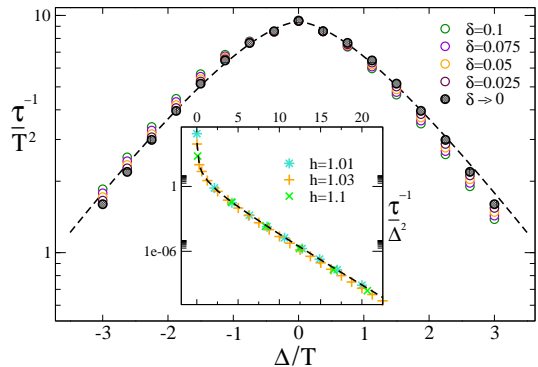}\\
\caption{$\tau^{-1}/T^{2}$ as a function of $\Delta/T$
(negative values correspond to $h<h_{c}$). Data are relative to different
distances $\delta=\sqrt{T^{2}+(h-h_{c})^{2}}$ from the QPT.
Close to the critical point ($\delta\rightarrow 0$) all curves collapse into a unique scaling function.
Dashed line is a fit $a(1+b{\Delta}/{T})\exp\left\{-{\Delta}/{T}\right\}$
with $a\simeq 9.4(6)$ and $b\simeq 0.9(3)$.
Inset shows the data collapse of $\tau^{-1}/\Delta^{2}$ as a function of $\Delta/T$
for different values of $h$: $\tau^{-1}\propto\exp\left\{ -\Delta/T\right\}$ for $T\ll\Delta$.
\protect }
\label{fig:tau+spectrumR}
\end{figure}

\begin{figure}
\includegraphics[scale=0.4]{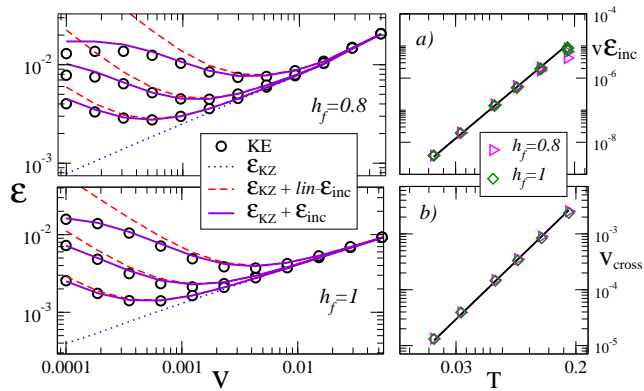}
\caption{Left: density of excitations $\mathcal{E}$ versus the quench velocity $v$.
Circles are obtained by integrating the kinetic equation (KE);
solid and dashed lines are Eq.~(\ref{eq:ebath+eclean}) with $\mathcal{E}_{inc}$
given by (\ref{eq:ebath_approx}) and its linearized expression, respectively.
KZ scaling $\mathcal{E}_{KZ}\propto \sqrt{v}$ \cite{zurek05} is plotted for comparison as dotted lines.
Plots refer to $N=400$ and $\alpha=0.01$ and $T=0.15,\ 0.1,\ 0.07$ (from top to bottom).
The final values of magnetic field are $h_{f}=0.8$ (outside the quantum critical region
for the values of $T$ considered) and the critical point $h_{f}=1$.
For $h_{f}=1$ we considered $t_{QC}=v^{-1}T$ since only half of the quantum
critical region is crossed.
a) Data collapse of $\mathcal{E}_{inc}$ obtained from the kinetic equations by considering
only incoherent thermal transitions (rhs of (\ref{eq:Kinetic}));
data refer to both $h_{f}=0.8,\ 1$ and $10^{-4}\lesssim v\lesssim 10^{-3}$.
The fit verifies the scaling (\ref{eq:Ebath_generic_QPT}) $\mathcal{E}_{inc}\propto v^{-1}T^{4}$.
In b) the scaling of $v_{cross}$ is obtained equating $\mathcal{E}_{inc}$
from kinetic equations to $\mathcal{E}_{KZ}$ for $h_{f}=0.8,\ 1$;
the fit confirms the scaling predicted by (\ref{eq:vbath_generic_QPT}),
$v_{cross}\propto T^{8/3}$.}
\label{fig:tau_approx}
\end{figure}

Eqs.~(\ref{eq:ebath+eclean}),(\ref{eq:Ebath_generic_QPT}),(\ref{eq:vbath_generic_QPT})
represent the generalization of the scaling laws given in
Refs.\cite{zurek05,polkovnikov05} to the case of open quantum
critical systems. They are in principle amenable of an experimental
verification and constitute the key result of this work.
In the case of a quantum Ising chain, which we discuss in detail in the second part of the Letter,
the previous expressions specialize as follows.
The time spent within the quantum critical region scales as $t_{QC}=2Tv^{-1}$, since $\nu=z=1$.
We consider an Ohmic bath that acts as a random external magnetic field on each site (see Eq.~(\ref{eq:Hamiltonian})).
We find that the relaxation time in the quantum critical region scales as $\tau^{-1}\simeq\alpha T^{2}$
(see Fig.~\ref{fig:tau+spectrumR}), {\it i.e.}, $\theta=2$, in agreement with a Fermi golden rule argument.
It follows that the contribution to the defect production induced by the bath and the
crossover velocity scale as $\mathcal{E}_{inc}\propto\alpha v^{-1}T^{4}$ and $v_{cross}\propto\alpha^{2/3}T^{8/3}$,
respectively (see Fig.~\ref{fig:tau_approx}).

\paragraph*{Quantum Ising Model and kinetic equations.}
To support Eqs.~(\ref{eq:Ebath_generic_QPT}) and
(\ref{eq:vbath_generic_QPT}) we now study the physics of a quantum quench
for a quantum Ising model coupled locally to a set of Ohmic baths.
The locality of the system-bath coupling causes
the breaking of the translational symmetry of the closed system,
hence allowing the discussion of the quench dynamics in the presence
of relaxation of all elementary excitations. Notice that no
qualitative features are expected to emerge in the case where the
baths are correlated over a finite distance because  (in the scaling limit)
the correlation length is the largest lengthscale in the problem, and
details of bath correlations over microscopic distances should not
matter. It is also important to observe that long time correlations
induced by the bath may change the universality class of the
transition~\cite{Werner07}. Despite the intrinsic interest of this
issue, which found recently application in the context of the
physics of cold atoms~\cite{footnote}, we will not consider it here.
Therefore,  we will further assume that the bosons have a non-zero
inverse lifetime $\gamma \ll T$ which provides a natural cutoff-time
for the bath correlation functions.

The model we consider is then defined by the Hamiltonian
\begin{equation} \label{eq:Hamiltonian}
H=-\frac{J}{2}\!\sum_{j}^{N}\!\left\{\sigma_{j}^{x}\sigma_{j+1}^{x}
         \!+\! [h(t)\!+\! X_{j}] \sigma_{j}^{z}\right\} \!+ H_{B} \;.
\end{equation}
It consists of a chain of $N$ spins ($\sigma^{x}$ and $\sigma^{z}$
are Pauli matrices) with an Ising interaction and subject to a
transverse magnetic field $h(t)$. The bath couples to $\sigma^{z}$,
with $X_{j}=\sum_{\beta} \lambda_{\beta}
(b_{\beta,j}^{\dagger}+b_{\beta,j})$, where
$b_{\beta,j}^{\dagger}$($b_{\beta,j}$) are the creation
(annihilation) operators for the bosonic bath modes coupled to the
j-th spin.
The bath Hamiltonian is $H_{B}=\!\sum_{j,\beta}\!\omega_{\beta}b_{\beta j}^{\dagger}b_{\beta j}$.
The system-bath coupling is chosen to have Ohmic spectral densities
$\sum_{\beta}\lambda_{\beta}^{2}\delta(\omega-\omega_{\beta})=2\alpha\omega\exp(-\omega/\omega_{c})$,
where $\omega_{c}$ is a high-energy cutoff~\cite{WeissBOOK,commentOhmic}.
In the case of no coupling to the bath ($\alpha=0$) the system has a QPT at $h_{c}=1$,
and for $h<h_{c}$ a spontaneous magnetization along $x$ appears.
The gap $\Delta=|h-h_{c}|$ induces at finite temperature a V-shaped
crossover phase diagram \cite{SachdevBOOK}, as sketched in Fig.~\ref{fig1}.

We now analyze the problem by deriving a quantum kinetic equation
which allows us to calculate the density of defects produced after
the quench. This procedure allows us to describe the effect of the
environment also in regimes which are beyond the applicability of
the scaling laws deduced above, where universality is not expected
to hold. In order to describe the dynamics of (\ref{eq:Hamiltonian})
it is first convenient to map the spins onto spinless fermions by
means of a Jordan-Wigner transformation. In momentum space, the
Hamiltonian (\ref{eq:Hamiltonian}) reads:
\begin{eqnarray}
H=\sum_{k>0}\Psi_{k}^{\dagger}\hat{\mathcal{H}}_{k}\Psi_{k}
+ \frac{1}{\sqrt{N}}\sum_{k,q}\Psi_{k}^{\dagger}\hat{\tau}^{z}\Psi_{k+q}X_{q} + H_{B} \;,
\end{eqnarray}
where $\hat{\mathcal{H}}_{k}=-[\cos{(k)}+h(t)]\hat{\tau}_{z}+\sin{(k)}\hat{\tau}_{y}$,
$\hat{\tau}$ are Pauli matrices in the Nambu space defined by the two-component fermion
$\Psi_{k}^{\dagger}=\left(\begin{array}{cc} c_{k}^{\dagger} & c_{-k}\end{array}\right)$.
When $X_{q}=0$, the subspaces of the different $k$ modes are decoupled, the Hamiltonian
is quadratic and can be diagonalized by a Bogoliubov transformation yielding
$H=\sum_{k>0}\Lambda_{k}(\eta_{k}^{\dagger}\eta_{k}-\eta_{-k}\eta_{-k}^{\dagger})$
with $\Lambda_{k}=\sqrt{1+2h\cos k+h^{2}}$.
The interaction with the bath causes a mixing of modes with different momenta.
We determine the density of excitations by deriving a quantum kinetic equation for the
Green's function using the Keldysh technique~\cite{HaugBOOK}. This kinetic equation will
be expressed in terms of the fermion lesser Green's function $\hat{G}^{<}$,
which is a $2\times2$ matrix in Nambu space with components
$-i[G_{k}^{<}(t,t)]_{i,j}\equiv\langle\Psi_{k,j}^{\dagger}(t)\Psi_{k,i}(t)\rangle$.
Using a self-consistent Born approximation for the bath-mediated scattering of k-modes,
valid for weak bath-system coupling $\alpha\ll 1$, and the Markov approximation
(justified by the assumption of a cutoff time for the bosonic modes), we get:
\begin{eqnarray} \label{eq:Kinetic}
\partial_{t}\hat{G}_{k}^{<} &+& i\left[\hat{\mathcal{H}}_{k},\,\hat{G}_{k}^{<}\right]=\\
\frac{1}{N}\sum_{q} \hat{\tau}^{z} (\hat{1}+i\hat{G}_{q}^{<}) \hat{D}_{qk} \hat{G}_{k}^{<}
                            &+& \hat{\tau}^{z} \hat{G}_{q}^{<} \hat{D}_{kq}^{\dagger} (\hat{1}+i\hat{G}_{k}^{<})
+ H.c. \nonumber
\end{eqnarray}
where we neglect irrelevant Lamb shifts. Here $\hat{D}_{qk}=i\int_{0}^{\infty}ds\,
g^{>}(s) {\hat{\mathcal U}}_{q}^{\dagger}(t,t-s) \hat{\tau}^{z} {\hat{\mathcal U}}_{k}(t,t-s)$,
where $g^{>}(t)=-i\left\langle X_{q}(t)X_{q}(0)\right\rangle$, and
${\hat{\mathcal U}}_{k}(t_{0},t)$ is the evolution operator satisfying
$i\partial_{t}{\hat{\mathcal U}}_{k}={\hat{\mathcal H}}_{k}(t) {\hat{\mathcal U}}_{k}$.
Parameterizing $-i\hat{G}_{k}^{<}=1/2[\hat{1}-(1-2P_{k})\hat{\tau}_{z}+C_{k}\hat{\tau}^{+}+C_{k}^{*}\hat{\tau}^{-}]$ after the Bogoliubov transformation which diagonalizes ${\hat{\mathcal H}}_{k}$,
one finds that $P_{k}=\langle\eta_{k}^{\dagger}\eta_{k}\rangle$ and
$C_{k}=\left\langle \eta_{-k}\eta_{k}\right\rangle$.
Therefore, the density of defects produced can be expressed as
\begin{equation}
\mathcal{E}  = \frac{-i}{2N}\sum_{k>0}
{\rm Tr} \left[ (\hat{1}+\hat{\tau}^z) \hat{G}_{k}^{<} \right] =\frac{1}{N}\sum_{k>0}P_{k}\;.
\label{eq:excitation_density}
\end{equation}
By solving the kinetic equation (\ref{eq:Kinetic}) we are able to calculate $\mathcal{E}$
in Eq.~(\ref{eq:excitation_density}), and thus to analyze the Kibble-Zurek mechanism in an Ising
chain coupled to a bath. The results of this analysis are presented below.

%

\paragraph{Kinetic equation and the scaling regime.}
%
%
As discussed in the first part of this Letter, in order to determine
the scaling at finite temperature we need to know the relaxation time $\tau$
for the excitations. To this end, it is sufficient to consider only
the dynamics of the populations $P_{k}$ in the kinetic equation and
neglect the off-diagonal components. By linearizing the kinetic
equation around the equilibrium (Fermi) distribution, one gets
$\partial_{t}{\delta P}=-\mathcal{R}(h,T){\delta P}$,
where ${\delta P}=\left(P_{k_{1}}-P_{k_{1}}^{th},\dots,P_{k_{N/2}}-P_{k_{N/2}}^{th}\right)$.
%
The characteristic relaxation times of the system are the inverse
eigenvalues of $\mathcal{R}$. The leading asymptotics at long times
for all populations $P_{k}$ is governed by the inverse of the smallest
eigenvalue $\lambda_{1}$ of $\mathcal{R}$, $\tau\equiv\lambda_{1}^{-1}$.
By numerical inspection, we find that  $\tau^{-1}\sim\alpha T^{2}f(\frac{\Delta}{T})\ e^{-\Delta/T}$
%
%
(see Fig \ref{fig:tau+spectrumR}).

The scaling obtained for the relaxation time, together with an explicit integration of
Eq.~(\ref{eq:ebath_approx_general}) for the Ising model, lead to
\begin{equation}
\mathcal{E}_{inc}\simeq\frac{\log2}{2\pi}T\left(1-e^{-2T/(\tau v)}\right) \;,
\label{eq:ebath_approx}
\end{equation}
which, in the limit $2T/\tau v\ll 1$, confirms the scaling result given in Eq.~(\ref{eq:Ebath_generic_QPT}).
In Fig.~\ref{fig:tau_approx} the solution of the kinetic equation
is compared with the  ansatz given in Eqs.~(\ref{eq:ebath+eclean}) and (\ref{eq:ebath_approx}).
The agreement is excellent, confirming our scaling approach.
The crossover value of $v$ which signals the transition from the coherent- to the incoherent-dominated
defect production obeys the power-law scaling given by Eq.~(\ref{eq:vbath_generic_QPT}).
At lower quench rates the full expression, Eq.~(\ref{eq:ebath_approx}),
is needed for an accurate comparison with the solution of the kinetic equation.

%
\begin{figure}
\includegraphics[scale=0.32]{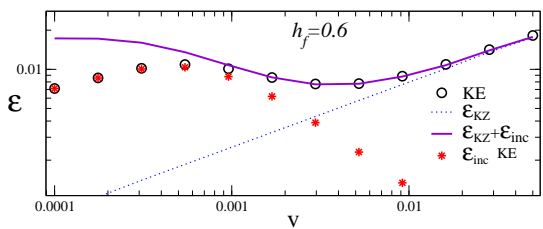}
\caption{Same as left panel of Fig. \ref{fig:tau_approx}, but for $h_f=0.6$;
curves are relative to $T=0.15$ only.
Stars represent the density of excitations produced
incoherently by the bath and are obtained by integrating
the kinetic equations and ignoring the unitary evolution term
$i\left[\hat{\mathcal{H}}_{k},\,\hat{G}_{k}^{<}\right]$,
responsible for the coherent excitation process.
The non-monotonous behavior is due to the non-critical relaxation
mechanism, that is more relevant at low $v$.
In the regime of slow $v$ scaling no longer holds and
only thermal excitations contribute to $\mathcal{E}$. }
\label{fig:fig5}
\end{figure}

On lowering the final value $h_f$ of the field, the agreement with
the scaling  ansatz becomes worse at low quench rates $v$ (see Fig.
\ref{fig:fig5}). This is due to a non-critical relaxation mechanism
that depends strongly on the details of the energy spectrum for
$h_{f}<h<h_{c}$: Once the system leaves the quantum critical region,
entering the semiclassical region where $T\ll\Delta$, the bath
starts to relax-out the excitations previously created; if time
spent in the semiclassical region is long enough all excitations
disappear. Outside of the weak coupling regime, when the
relaxation time and quench time are comparable, scaling is expected
to be recovered in thermodynamic quantities, such as total energy,
heat, entropy and work done on the system~\cite{thermo}.

We acknowledge F. Guinea, A.J. Leggett and F. Sols for fruitful discussions
and A. Polkovnikov and C. E. Creffield for their comments to the manuscript.
We acknowledge ESF (Exchange Grant 1758, INSTANS) and EU (Eurosqip and RTNNANO) for
financial support


\begin{thebibliography}{10}
\bibitem{commentOhmic}  For non-Ohmic spectral densities the scenario
in Fig (\ref{fig:tau_approx}) is preserved with a
modified exponent $\theta$.
\end{thebibliography}
\end{document}